\documentclass[12pt]{article}

\usepackage{amsmath,amsfonts,latexsym}
\usepackage{fullpage}

\def\01{\{0,1\}}
\newcommand{\ceil}[1]{\lceil{#1}\rceil}

\newcommand{\ket}[1]{|#1\rangle}

\newcommand{\Tr}{\mbox{\rm Tr}}

\newtheorem{theorem}{Theorem}
\newtheorem{lemma}{Lemma}

\newcommand{\COMMENT}[1]{}
\newenvironment{proof}
{\noindent {\bf Proof. }}
{{\hfill $\Box$}\\
 \smallskip}

\begin{document}

\title{Quantum multiparty communication complexity and circuit lower bounds}
\author{Iordanis Kerenidis\\MIT}
\maketitle

\begin{abstract}

We define a
quantum model for multiparty communication complexity and prove a
simulation theorem between the classical and quantum models. As a
result of our simulation, we show that if the
quantum $k$-party communication complexity of a function $f$ is
$\Omega(\frac{n}{2^k})$, then its classical $k$-party communication is
$\Omega(\frac{n}{2^{k/2}})$. Finding such an $f$ would allow us to
prove strong classical lower bounds for $k\geq \log n$ players and
hence resolve a main open question about symmetric
circuits. Furthermore, we prove that for the Generalized Inner Product
($GIP$) function, the quantum model is exponentially more efficient
than the classical one. This provides the first exponential separation
for a total function between any quantum and public coin randomized
communication model.  

\end{abstract}

%--------------------------------------------------------------------

\section{Introduction}

Communication complexity is a central model of computation with 
numerous applications. It has been used for proving lower bounds in 
many areas including Boolean circuits, time-space tradeoffs,
data structures, automata, formulae size, etc. Examples of these
applications can be found in the textbook of Kushilevitz and Nisan
\cite{KN97}.  

The ``Number on the Forehead'' model of multiparty communication
complexity was 
introduced by Chandra, Furst and Lipton \cite{CFL}. In this model,
there are $k$ parties that wish to compute a function $f:X_1\times
\cdots \times X_k \rightarrow \{0,1\}$ on the input
$(x_1,\ldots,x_k)\in (X_1\times\cdots \times X_k)$. We can assume that
$X_1=\ldots = X_k=\{0,1\}^n$. Each player can see only $(k-1)$ of the
inputs (the other one is on his forehead). The players communicate by
writing messages on a common blackboard that they can all see. In each
round, one player writes one bit in the blackboard. At the end of the
protocol, the blackboard must contain enough information to compute
the value of $f(x_1,\ldots,x_k)$. The communication cost of the
protocol is the number of bits written on the blackboard. The
deterministic k-party communication complexity of $f$, $C(f)$, is the
communication cost of the optimal deterministic protocol for $f$. In
the randomized setting, we allow the players to be probabilistic and
the output of the protocol to be correct with probability at least
$1/2+\delta$. We define $C_\delta (f)$ to be the probabilistic k-party
communication complexity of $f$ with correctness $1/2+\delta$.

In the above definition the number of players was equal to the number
of arguments of $f$. However, we can easily generalize the model for
the case of $\ell \leq k$ players. The model of communication remains
the same and each of the $\ell$ players still receives $(k-1)$
arguments of $f$. We denote 
with $C_\delta^{\ell}(f)$ the $\ell$-party communication complexity
of $f(X_1,\ldots,X_k)$.  

In the Simultaneous Messages variant of the Number on the Forehead,
all the players write simultaneously one message on the blackboard,
after which the value of $f$ can be computed with high
probability. The communication cost is the sum of the lengths
of the messages. 

Multiparty communication complexity has been studied extensively and
has proved relevant to important questions in circuit lower
bounds. For example, one of the major open problems in circuit
complexity is to prove that a function $f$ is not in the circuit
complexity class $ACC$. By the results of \cite{HG,Yao}, this reduces
to proving a superlogarithmic communication lower bound for the
$k$-party communication complexity of some explicit function $f$,
where the number of players is superlogarithmic. 
All known techniques for proving multiparty communication lower
bounds fail when the number of players becomes $k=\log n$. 

In this
paper, we propose a new technique for proving multiparty communication
complexity lower bounds and hence, circuit lower bounds. We define a
quantum model for multiparty communication complexity and prove a
simulation theorem between the classical and quantum models. More
specifically, we show how to simulate $k$ classical players with only
$k/2$ quantum ones.
This enables us to reduce questions about classical communication
to potentially easier questions about quantum communication complexity
and shows that quantum information theory could be a
powerful tool for proving classical circuit lower bounds. 

Such connections between classical
and quantum computation have been proved to be very fruitful in the
last few years. Important results in classical complexity theory were proved
using quantum techniques or inspired by them, for example lower bounds
for Locally Decodable Codes \cite{KW03} or local search \cite{Aar1},
inclusions of lattice problems in complexity classes \cite{AR1,AR2},
simple proofs of properties of the class PP \cite{Aar2}. 

In addition, we examine the power of
quantum multiparty communication complexity and show that for the
Generalized Inner Product ($GIP$) function, the quantum model is
exponentially more efficient than the classical one. This provides the
first exponential separation for a total function between any quantum and
public coin randomized communication model. 
Exponential separations in the 2-party setting have been proved in the
models of two-way communication \cite{Raz2}, one-way communication
\cite{BJK} and Simultaneous Messages \cite{BJK}. Note that all these
separations are for promise problems or relations and not for total
boolean functions.

\subsection{Multiparty communication complexity and circuit lower bounds}

Multiparty communication complexity was introduced as a tool for the
study of boolean circuits, however the known techniques for 
proving lower bounds are very limited. Babai et al. \cite{BNS} proved
a lower bound of $\Omega(\frac{n}{2^{2k}}+\log \delta)$ for the $k$-party
communication complexity of the Generalized Inner product function and
Chung \cite{Chung} improved it to $\Omega(\frac{n}{2^k}+\log
\delta)$. Raz \cite{Raz} simplified their proof technique and showed a
similar lower bound for another function, i.e. Matrix Multiplication,
which seems to be hard even for $\log n$ players. Unfortunately, the
above techniques are limited and cannot prove lower bounds 
better than $\Omega(\frac{n}{2^k}+\log \delta)$ for any
function. Despite the importance of the question and its serious
consequences on circuit lower bounds, it has not been possible to find
any new lower bound techniques. For the Generalized Inner Product
function, Grolmusz \cite{Gr} showed a matching upper bound of
$O(\frac{n}{2^k}+\log \delta)$.  

The Number on the Forehead model is related to the circuit
complexity class $ACC^0$. $ACC^0$ are constant-depth polynomial size,
unbounded fan-in circuits with $NOT,AND,OR$ and $MOD_m$ gates. It is a
major open question to find an explicit function outside the class
$ACC^0$. Yao \cite{Yao} and Beigel,Tarui \cite{BT} have shown that
$ACC^0$ circuits can be simulated by symmetric circuits. The circuit
class $SYM(d,s)$ is the class of circuits of depth 2, whose top gate
is a symmetric gate of fan-in $s$ 
and each of the bottom level gates is an AND gate of fan-in at most
$d$. Specifically, they showed that $ACC^0\subseteq SYM(polylog n,
2^{polylog n})$. The connection to multiparty communication was made
by Hastad and Goldmann \cite{HG}, who noticed that when a function $f$
belongs to $SYM(d,s)$, then there exists a $(d+1)$-party simultaneous
protocol with complexity $O(d \log s)$. Hence, if we want to show that
a function $f$ is outside $SYM(d,s)$, then we need to prove a
$(d+1)$-party communication lower bound of $\omega(d \log s)$. 
However, as we said, no techniques are known to give communication
lower bounds for $k=\log n$ players or more. In the next sections we
will describe a technique that can potentially give strong lower
bounds for $k\geq\log n$ players. This would be a first step
towards proving that a function is outside $ACC^0$ (see \cite{KN97},
Open problem 6.21). 

\subsection{Quantum background}

Let $H$ denote a 2-dimensional Hilbert space and $\{\ket{0},\ket{1}\}$
an orthonormal basis for this space.
A \emph{qubit} is a unit length vector in this space, and so can be
expressed as a linear combination of the basis states:
$ \alpha_0\ket{0}+\alpha_1\ket{1} $.
Here $\alpha_0,\alpha_1$ are complex \emph{amplitudes}
and $|\alpha_0|^2+|\alpha_1|^2=1$. 
An \emph{$m$-qubit system} is a unit vector in the $m$-fold tensor space
$H\otimes\cdots\otimes H$ and can be expressed as 
$ \ket{\phi}=\sum_{i\in\01^m}\alpha_i\ket{i}.$
A \emph{mixed state} $\{p_i,\ket{\phi_i}\}$ is a classical distribution
over pure quantum states, where the system is in state $\ket{\phi_i}$
with probability $p_i$. 

A quantum state can evolve by a unitary operation or by a
measurement. A \emph{unitary} transformation is a linear mapping that
preserves the $\ell_2$ norm. If we apply a unitary $U$ to a state
$\ket{\phi}$, it evolves to $U\ket{\phi}$.
A mixed state $\rho$ evolves to $U\rho U^{\dag}$.
The most general measurement (POVM) allowed by quantum mechanics is 
specified by a family of positive semidefinite operators 
$E_i=M_i^*M_i$, $1\leq i\leq k$, subject to the condition
that $\sum_i E_i=I$. Given a mixed state $\rho$, the probability of
observing the $i$th outcome under this measurement is given by the
trace $p_i=\Tr(E_i\rho)=\Tr(M_i\rho M_i^*)$. If the measurement yields
outcome $i$, then the resulting quantum state is $M_i\rho
M_i^*/\Tr(M_i\rho M_i^*)$. A general POVM can be thought of as a
series of unitary operations and projective measurements.

%------------------------------------------------------------

\section{Quantum multiparty communication complexity}

In order to make the definition of the quantum analog  more intuitive,
we are going to describe the classical model of the Number on the
Forehead in a different but equivalent way. 
Let us assume that $\ell$ players want to compute a function
$f:X_1\times \cdots \times X_k \rightarrow \{0,1\}$ on the input
$(x_1,\ldots,x_k)\in (X_1\times\cdots \times X_k)$. Without loss of
generality $X_1=\ldots = X_k=\{0,1\}^n$.  
 
The protocol is performed by a Referee and the $\ell$
players. The notion of the referee is mainly conceptual and it will be
clear that it doesn't change the power of the classical model at
all. Without loss of generality, we assume the $\ell$ players are
equivalent and their answers have the same size.\footnote{We can
achieve that by having each player play the role of each one of the
$\ell$ players and hence increase the communication by a factor of
$\ell$.} 
The protocol consists of three rounds and the communication
is done by writing on a blackboard. In fact, we can assume that
there are $\ell$ disjoint blackboards. Player $i$ can read and write
only on the $i$-th blackboard, though the referee can read and write
on all of them. \\

\noindent
{\bf Classical Number on the Forehead}
\begin{itemize}
\item 
In the first round, the referee writes on the $i$-th
blackboard a string $P_i$, which is the input to the
$i$-th player. The only valid inputs $P_i$ are strings that consist of
$(k-1)$ of the $x_i$'s. Without loss of generality, $P_j =
(x_1,\ldots,x_{j-1},x_{i+1},\ldots,x_k)$. 
\COMMENT{
The referee knows the input
$(x_1,\ldots,x_k)$ and so he can construct the inputs $P_i$. }
\item
In the second round, player $i$ reads input $P_i$ and writes
his answer $A_i$ on the blackboard.  
\item
In the third round, the blackboard contains the strings
$(P_1,A_1),\ldots , (P_\ell,A_\ell)$. 
First the referee ``erases'' the inputs $P_i$ and then computes
$g(A_1,\ldots,A_\ell)$ as his guess for $f(x_1,\ldots,x_k)$. The
function $g$ is fixed by the protocol and is independent of the inputs
$(x_1,\ldots,x_k)$.    
\end{itemize}

The correctness of the protocol says that for every
$(x_1,\ldots,x_k)\in \{0,1\}^{kn}$ we have $Pr[g(A_1,\ldots,A_\ell)~=~
f(x_1,\ldots,x_k)]\geq 1/2+\delta$. The communication cost of the
protocol is the sum of the lengths of the messages that the players
write on the blackboard, i.e. $\sum_{i=1}^\ell |A_i|$ and the
communication complexity of $f$ is the cost of the optimal protocol. 
It's easy to see that the model described above is equivalent to
the usual Number on the Forehead model. In addition, it
makes the definition of the quantum analog more intuitive. 

Our goal is to define a quantum model for multiparty communication,
which is powerful enough to be interesting, but simple enough to
facilitate the proof of strong lower bounds. In this model, 
we allow the referee to create quantum inputs for the players,
but we ensure that each quantum player obtains information for at
most $(k-1)$ of the inputs $x_i$. The players read these inputs and
write their answers on the blackboards. The referee, then, quantumly
``erases'' the input states and performs a general measurement (POVM)
on the answers. The outcome of the measurement is his guess for the
value of $f(x_1,\ldots,x_k)$.    

A quantum
``blackboard'' is a Hilbert space that consists of three parts
$\cal{H}_A\otimes\cal{H}_B\otimes\cal{H}_C$, where $\cal{H}_A$ is the
workspace of the referee, $\cal{H}_B$ is the space where the referee
writes the input to the player and $\cal{H}_C$ is the space where the
player writes his answer. We assume that we have $\ell$ such
blackboards which are disjoint (unentangled). More formally, the
quantum model is defined as follows: \\\\\\

\noindent
{\bf Quantum Number on the Forehead}
\begin{itemize}
\item
In the first round, the referee constructs the inputs of the
players. The only valid input to quantum player $i$ is a mixed
state of the form $\rho_i = \{p^i_j, P_j\}$, i.e. a distribution over
the valid classical inputs $\{P_j\}$. The distribution is fixed by the
protocol and is independent of the input $(x_1,\ldots,x_k)$. Without
loss of generality, the referee performs the following operations for 
$i=1,\ldots,\ell$:\\ 
First he constructs a state $\sum_{j=1}^k \sqrt{p^i_j}\ket{j}_{A_i}$ 
and then performs a mapping $T$
\[ T: \;\ket{j}_{A_i}\ket{0}_{B_i} \mapsto \ket{j}_{A_i}\ket{P_j}_{B_i} \]
resulting in the state 
\[ \ket{\phi_i} = \sum_{j=1}^k\sqrt{p^i_j}\ket{j}_{A_i}\ket{P_j}_{B_i}.\]
The register $A_i$ is the referee's workspace and has size $\log
k$. The register $B_i$ contains the input to player $i$.   
\item
In the second round, player $i$ ``reads'' his input from register
$B_i$ and ``writes'' his answer on register $C_i$. 
Specifically, quantum player $i$ performs the following mapping: 
\[ \ket{P_j}_{B_i}\ket{0}_{C_i} \mapsto
\ket{P_j}_{B_i}\ket{Q^i_j}_{C_i}\]   
\item
In the third round, blackboard $i$ contains the state 
\[ \ket{\psi_i} =
\sum_{j=1}^k\sqrt{p^i_j}\ket{j}_{A_i}\ket{P_j}_{B_i}\ket{Q^i_j}_{C_i}.\] 
The referee ``erases'' the inputs $P_j$ by performing the
inverse mapping $T^{-1}$, i.e.  
\[ T^{-1}:\; \ket{j}\ket{P_j} \mapsto \ket{j}\ket{0},  \]
resulting in the states
\[\ket{\psi_i}=\sum_{j=1}^k\sqrt{p^i_j}\ket{j}_{A_i}\ket{Q^i_j}_{C_i}.\]
Then, he performs a general measurement $M$ on these states, whose
outcome is his guess for $f(x_1,\ldots,x_k)$. The quantum procedure
$M$ is fixed by the protocol and is independent of the input $x$.\\    
\end{itemize}

\noindent
The correctness of the protocol guarantees that for every
$(x_1,\ldots,x_k)\in \{0,1\}^{kn}$, 
\[Pr[\mbox{outcome of }M = f(x_1,\ldots,x_k)]\geq
1/2+\delta.\] The communication cost of the 
protocol is the sum of the lengths of the messages that the players
write on the blackboard, i.e. $\sum_{i=1}^\ell |C_i|$, where $|C_i|$ is
the size of the answer register of player $i$.\footnote{More
precisely, the communication should be defined as 
$\sum_{i=1}^\ell (|A_i|+|C_i|)$, however the communication according
to this definition is in the worst case an additive factor of $\ell
\log k$ greater than our definition which will not be of any significance.}  
The communication complexity of $f$ is the cost of the optimal
protocol.\\ 

\noindent
{\bf Remarks: \\}
{\bf 1)} The inputs $\{p^i_j,P_j\}$ are the only quantum inputs which ensure
that each player gains information for at most $(k-1)$ of the
inputs $x_i$. For example, an input of the form $\sum_{j=1}^k
\ket{P_j}$ enables the quantum player to learn $(k-1)$ arbitrary bits
of information about $(x_1,\ldots,x_k)$. \\
{\bf 2)} A naive method of ``erasing'' in the quantum case would be to just
ignore the second register. However, it can be shown that this is
equivalent in having the inputs being classical strings and hence this
model is not very appealing.

%--------------------------------------------------------------

\section{Simulating classical players}
\label{simulation} 

In the previous section we defined a quantum model for multiparty
communication. We will prove that in this model we can simulate a
$k$-party classical protocol by a $k/2$-party quantum protocol with
the same communication, albeit with larger error probability.

\begin{theorem}\label{sim}
Let $P$ be a multiparty protocol for the function
$f:X_1,\ldots,X_k\rightarrow \{0,1\}$
with $k$ players, communication $C$ and correctness
$1/2 + \delta$. Then there exists a quantum protocol $Q$ for the same
function $f$ with $k/2$ quantum players, communication
$C$ and error $1/2 + \delta/2^{3C/2}$ on an average input. 
\end{theorem}

\begin{proof}
First, we prove a lemma similar to Lemma 2 in \cite{KW03}, which shows
that we can assume the referee computes the parity of a subset of the
answer bits as his guess for $f$. 

\begin{lemma}
let $P$ be a classical protocol with communication $C$, where the
referee computes a function $g(A_1,\ldots,A_k)$ as his guess for
$f(x_1,\ldots,x_k)$, where $A_i$ is the answer of player $i$. Then,
there exists a classical protocol $P'$ with 
communication $C$ that works on average input with correctness
$1/2 + \delta/2^C$ and where the referee computes a parity of a
subset of bits of the answers $A_i$, i.e.$g(A_1,\ldots,A_k) =
\oplus S_i$, where $S_i$ denotes the parity of a subset of bits of
$A_i$.    
\end{lemma}

\begin{proof}(Lemma)
Let $f(x_1,\ldots,x_k)=b$ and $x = x_1,\ldots,x_k$.
From the correctness of the protocol $P$ we know that
$E_x[g(a_1,\ldots,a_k)\cdot b] \geq 2\delta $. Using the Fourier
representation of $g$ we have
\[ 2\delta \leq E_x[g(a_1,\ldots,a_k)\cdot b] = \sum_{S_1,\ldots,S_k}
\hat{g}_{S_1,\ldots,S_k} E_x[\prod_{S_1} a_{S_1}\cdots \prod_{S_k}
a_{S_k}\cdot b] 
\]
Averaging and using the fact that $|\hat{g}|\leq 1$ we get that there
exist some subsets $S_1,\ldots,S_k$ for which
\[E_x[\prod_{S_1} a_{S_1}\cdots \prod_{S_k} a_{S_k}\cdot b]\geq
2\delta / 2^C. 
\]
This means that the protocol $P'$ which would output the XOR of these
subsets is correct on an average input with probability $\leq 1/2+
\delta / 2^C$. 
\end{proof}

Hence, in the classical protocol $P'$, the referee gives input $P_i$ to
player $i$, the players write the answers $A_i$ and the referee computes
$f$ by taking the $XOR$ of a subset of the bits of the $A_i$'s. 
Now we will describe the quantum protocol with only $k/2$ players that
simulates the classical $k$-party one. We denote the $k/2$ quantum
players with $i=1,3,\ldots,k-1$.

\begin{itemize}
\item
In the first round, the referee creates the following states: 
\[ \ket{\phi_i} =
\ket{i}_{A_i}\ket{P_i}_{B_i}+\ket{i+1}_{A_i}\ket{P_{i+1}}_{B_i},\] 
where the second register is the input of quantum player $i$ and the
first one is the purification of the state in the referee's
workspace. Note that the reduced  
density matrix of quantum player $i$ is the same as if he was
classical player $i$ with probability $1/2$ and classical player $i+1$
with probability $1/2$. Hence, this is a legal input. 
\item
In the second round, each player performs the following
mapping: 
\[  T: \ket{P_j}\ket{0} \mapsto \ket{P_j}\ket{A_j}, \]
i.e. on input $\ket{P_j}$ computes the same function $A_j$ as the
classical player $j$ in $P$.  
\item
In the third round, the ``blackboard'' contains the states
\[ \ket{\phi_i} =
\ket{i}\ket{P_i}\ket{A_i}+\ket{i+1}\ket{P_{i+1}}\ket{A_{i+1}}  \]
The referee ``erases'' the input register resulting in the states  
\[ \ket{\psi_i} = \ket{i}\ket{A_i}+\ket{i+1}\ket{A_{i+1}}.\]
Last, the referee performs a measurement on these states (described by
Lemma 2) and computes $f$ with high probability. 
\end{itemize}

We need to show that there exists a quantum procedure $M$ on the
states $\ket{\psi_i}$ that enables the referee to compute the function
$\oplus S_i$. A key observation is that we can rewrite the function as
\[ \oplus_i  S_i = \oplus_{i=1,3,\ldots,k-1}(S_i \oplus S_{i+1}). \] 
It's a simple calculation to show that if we can predict $S_i\oplus
S_{i+1}$ with probability $1/2+\epsilon$ then we can predict the
entire $\oplus_i  S_i$ with probability $1/2+2^{k/2-1}\epsilon^{k/2}$.   
The following lemma from \cite{WW04} describes a
quantum procedure $M$ to compute $S_i\oplus S_{i+1}$ with the optimal
$\epsilon$.   

\begin{lemma} (Theorem 2,\cite{WW04})
Suppose $f:\{0,1\}^{2t}\rightarrow \{0,1\}$ is a boolean
function. There exists a quantum procedure $M$ to compute $f(a_0,a_1)$
with success probability $1/2+1/2^{t+1}$ using only one copy of
$\ket{0}\ket{a_0}+\ket{1}\ket{a_1}$, with $a_0,a_1 \in \{0,1\}^t$.
\end{lemma} 
We use this lemma with $t=C/k$ and get $\epsilon=1/2^{C/k+1}$. Hence,
there exists a quantum procedure that will output 
the correct $\oplus_i S_i$ with probability
\[ Pr[ M \mbox{ outputs } \oplus_i S_i]=
1/2+2^{k/2-1}\cdot \frac{1}{2^{(C+k)/2}} = \frac{1}{2}+ \frac{1}{2^{C/2+1}}.\]
Finally, the quantum protocol is correct with probability
\begin{eqnarray*} 
p & = & Pr[M \mbox{ outputs }\oplus S_i]\cdot Pr[\oplus S_i = b]+
Pr[M \mbox{ doesn't output } \oplus S_i]\cdot Pr[\oplus S_i \neq b] \\
& = &(\frac{1}{2}+\frac{1}{2^{C/2+1}})(\frac{1}{2}+\frac{\delta}{2^C})+
(\frac{1}{2}-\frac{1}{2^{C/2+1}})(\frac{1}{2}-\frac{\delta}{2^C}) =
\frac{1}{2}+ \frac{1}{2^{3C/2}}  
\end{eqnarray*}
\end{proof}

%---------------------------------------------------------------

\section{A quantum reduction for circuit lower bounds}

The theorem in the previous section shows how to simulate a classical
protocol with $k$ players with a quantum protocol with $k/2$
players. We are going to use this theorem in order to get a reduction  
from a classical circuit lower bound question to one about quantum
communication complexity. 

\begin{theorem}
Suppose $f:X_1\times\cdots\times X_k \rightarrow \{0,1\}$ is a
function for which the $(\frac{k}{2})$-party quantum communication
complexity is $Q^{k/2}_\delta = \Omega(\frac{n}{2^{k/2}}+\log
\delta)$. Then this function does not belong to the class
$SYM(k-1, 2^{o(n/2^{k/2})})$. 
\end{theorem} 

\begin{proof}
Let the function $f$ have $(\frac{k}{2})$-party quantum communication
complexity 
$QC^{k/2}_\delta \geq \gamma ( \frac{n}{2^{k/2}} + \log\delta),$ for a
positive constant $\gamma$.   
Assume that the classical $k$-party communication complexity is
$C_\delta \leq \frac{\gamma}{2+3\gamma/2}
(\frac{n}{2^{k/2}}+\log \delta)$, then    
by Theorem \ref{sim} there exists an $(\frac{k}{2})$-party quantum
protocol with correctness $1/2 + \delta/2^{3C_\delta/2}$ and quantum
communication  $QC^{k/2}_{\delta/2^{3C_\delta/2}} = C_\delta$.
This contradicts the lower bound on $QC$ since 
\[ QC^{k/2}_{\delta/2^{3C_\delta/2}} 
\geq  \gamma ( \frac{n}{2^{k/2}}+\log \frac{\delta}{2^{3C_\delta/2}}) 
% =  \gamma (  \frac{n}{2^{k/2}} -3C_\delta/2 + \log \delta) 
\geq
2\left(\frac{\gamma}{2+3\gamma/2}(\frac{n}{2^{k/2}}+\log\delta)\right)
> C_\delta \]
By \cite{HG} the function $f$ does not belong in the class
$SYM(k-1,2^{o(n/2^{k/2})})$.   
\end{proof}

Taking $k=\log n + 1$ implies that the function $f$ is not in
$SYM(\log n, 2^{o(\sqrt{n})})$. In other words, we reduced the
question of finding a function outside the class $SYM(\log n,
2^{\omega(polylog n)})$ to that of finding an explicit function
$f:X_1\times\cdots\times X_k \rightarrow \{0,1\}$ with
$(\frac{k}{2})$-party quantum communication complexity equal to 
$\Omega(\frac{n}{2^{k/2}}+\log\delta)$. Note that we do know explicit
functions for which the classical communication is exactly of this
form, e.g. the functions $GIP$ (\cite{Chung}) and Matrix   
Multiplication (\cite{Raz}). In fact, the proofs given in these papers
consider only $k$-party communication, but as we'll see in section
\ref{randLB} they can easily be modified for the case of $\ell \leq k$
parties. We believe that quantum communication complexity can be a
very powerful tool for proving circuit lower bounds beyond the known
classical techniques.

%------------------------------------------------------------------

\section{An exponential separation}
\label{exp}
 
In this section we prove an exponential separation between the
classical and quantum multiparty communication complexity
model. Separations 
between classical and quantum two party communication models have been
found before, e.g. in the two-way model \cite{Raz}, one-way
\cite{BJK}, Simultaneous Messages \cite{BJK}. These separations are
for promise problems or relations and not for total boolean functions. Our  
separation on the other hand is the first one for a total boolean
function, namely the Generalized Inner Product function. Let us note
that Burhman {\em et al} \cite{BCWW01} showed an exponential
separation in the two party Simultaneous Messages model for a total
function, however that separation does not hold if we allow the
classical players to share public coins. \\

\noindent
{\bf The Generalized Inner Product Function $GIP(X_1,\ldots,X_k)$}\\

Let $X_i\in\{0,1\}^n$. We can think of the $k$ inputs as the rows of a
$k\times n$ matrix. Then $GIP(X_1,\ldots,X_K)$ is equal to the number
$(\mod 2)$ of the columns of the matrix that have all elements equal to 1.
More formally, denote with $X_i^j$ the $(i,j)$ element of this
matrix (which is equal to the $j$-th bit of $X_i$), then
\[ GIP(X_1,\ldots,X_k) = \;\;\sum_{j=1}^{n} \prod_{i=1}^k X_i^j  (\mod
2) \] 

The function $GIP$ has been studied extensively in the multiparty
communication model. Babai {\em et al.} \cite{BNS} showed a
$\Omega(\frac{n}{2^{2k}})$ lower bound and Chung \cite{Chung} improved
it to $\Omega(\frac{n}{2^{k}})$. The lower bound holds in the general
multiparty model where the answers of the players may depend on
previous answers. Moreover Grolmusz \cite{Gr} showed a
matching upper bound of $O(\frac{n}{2^{k}})$. We are going to use
these tight bounds and our quantum reduction in Theorem \ref{sim} to
prove an exponential separation between the quantum and randomized
$\ell$-party communication complexity of $GIP(X_1,\ldots,X_k)$.

\subsection{The randomized communication complexity of $GIP$}
\label{randLB}

\begin{theorem}
\label{rlb}
The $\ell$-party randomized communication complexity of
$GIP(X_1,\ldots,X_k)$ is 
\[  C_\delta^{\ell}(GIP) = \Omega(\frac{n}{2^\ell}+\log \delta) \]
For $k=\log (n+1)+1$, $\ell = \ceil{\frac{k}{2}}$ and any
constant $\delta$, we have $C_\delta(GIP) = \Omega(\sqrt{n})$.   
\end{theorem}

\begin{proof}
Assume that there exists an $\ell$-party communication protocol $P$ for
the function $GIP(X_1,\ldots,X_k)$ with communication $C_\delta^\ell$
and correctness $\frac{1}{2}+\delta$. Without loss of generality the 
input to the player $i$ is  
$P_i=(x_1,\ldots,x_{i-1},x_{i+1},\ldots,x_k)$.  
This implies
an $\ell$-party protocol $P'$ for the function
$GIP(X_1,\ldots,X_\ell)$ 
with the same communication $C_\delta^\ell$ and correctness
probability $\frac{1}{2}+\delta$. The players upon receiving input 
$P'_i~=~(x_1,\ldots,x_{i-1},x_{i+1},\ldots,x_\ell)$ they transform
them into the $k$-argument inputs 
$(x_1,\ldots,x_{i-1},x_{i+1},\dots,x_\ell,{\bf
1},\ldots,{\bf 1})$ and 
they execute protocol $P$. It's easy to see that
$GIP(x_1,\ldots,x_\ell,{\bf 1},\ldots,{\bf 1})=GIP(x_1,\ldots,x_\ell)$
and hence the protocol $P'$ has the same correctness probability
$\frac{1}{2}+\delta$ and the same communication $C_\delta^\ell$.
Chung's (\cite{Chung}) lower bound for the $\ell$-party communication
complexity of $GIP(X_1,\ldots,X_\ell)$ implies that
$C_\delta^\ell=\Omega(\frac{n}{2^\ell}+\log \delta)$.  
\end{proof}

As we said, Chung \cite{Chung} described a general method for
proving lower bounds up to $\Omega(\frac{n}{2^k})$ for the
$k$-party communication complexity of functions and explicitly shown
such a bound for $GIP(X_1,\ldots,X_k)$. A conceptually easier proof of
the same results was given by Raz \cite{Raz}. 
We can get an alternative proof of our Theorem \ref{rlb} by modifying
Raz's technique for the case of $\ell$-party communication
complexity. All these bounds hold in the general multiparty 
model and not just in the Simultaneous messages. The lower bound of
$\Omega(\sqrt{n})$ holds even when the protocol is 
correct only on an average input and the correctness probability is
$\frac{1}{2}+\frac{1}{poly(n)}$.

%------------------------------------------------------------------

\subsection{The quantum communication complexity of GIP}

Grolmusz \cite{Gr} described a $k$-party communication protocol for
$GIP(X_1,\ldots,X_k)$ with communication
$(2k-1)\ceil{\frac{n}{2^{k-1}-1}}$. 
Using our simulation from Theorem \ref{sim}, we can show
that there exists a quantum $(\frac{k}{2})$-party communication
protocol for $GIP$ with the same communication that is correct on
average with probability $\frac{1}{2}+\frac{1}{poly(n)}$. For $k= \log
(n+1)+1$ the quantum communication is only $O(\log n)$. This
already establishes the exponential separation, since as we said, the
classical lower bound for $k= \log (n+1)+1$ parties and average
correctness $\frac{1}{2}+\frac{1}{poly(n)}$ is $\Omega(\sqrt{n})$.  

For the specific $GIP$ function we can provide a better
simulation and show an efficient quantum algorithm which is correct on
all inputs with probability $1/2+\delta$. As we will see, this quantum
protocol is not simultaneous, but slightly more general. We avoid
redefining in full generality the quantum multiparty model for
non-simultaneous messages, first because lower bounds for the quantum
simultaneous messages model are sufficient for our reductions about
circuits,  and second, because the protocol we will present is
non-simultaneous in a very simple way. Note that the classical lower
bounds hold for the most general multiparty communication model.

\begin{theorem}
Let $k= \log (n+1)+1$, $\ell=\ceil{k/2}$ and any constant
$\delta$. Then, the $\ell$-party quantum communication complexity of 
$GIP(X_1,\ldots,X_k)$ is $ QC^\ell_\delta(GIP) = O(\log n)$.

\end{theorem}

\begin{proof}
As we mentioned, Grolmusz \cite{Gr} showed a $k$-party protocol
for $GIP(X_1,\ldots,X_k)$ with communication
$(2k-1)\ceil{\frac{n}{2^{k-1}-1}}$. Taking $k= \log (n+1)+1$ the
communication cost is $(2k-1)$ bits. In fact, the first player
communicates a $(k-1)$-bit string and a single bit and the other
$(k-1)$ players communicate a single bit each. The final answer is the
Parity of the single bits. The
single bits of the $(k-1)$ players depend on the message of the first
player and hence this is not a simultaneous messages
protocol.\footnote{We can transform this protocol into a simultaneous 
messages one by having each player send $O(k)$ bits.} Let us also
assume without loss of generality that $k$ is odd. 
We are going to simulate exactly the protocol of Grolmusz by using only
$\ceil{\frac{k}{2}}$ quantum players. \\

\noindent
{\bf Quantum protocol}\\

Let $P_1,\ldots,P_k$ be the inputs to the $k$ players in Grolmusz's
protocol and $A_1,\ldots,A_k$ the messages they write on the
blackboard. As we said, $A_1 = \in\{0,1\}^{k-1}\times\{0,1\}$ and for
$i=2,\ldots,k$ $A_i$ is a bit that depends on $(P_i,A_1)$. The idea is
to use the first quantum player to simulate exactly the first
classical player and for the other players use our simulation
technique form section \ref{simulation}. More specifically,

\begin{itemize}
\item
In the first round, the referee creates the following states:
\[ \ket{\phi_1}=\ket{P_1},\;\;\; \ket{\phi_i} =
\ket{i}\ket{P_i}+\ket{i+1}\ket{P_{i+1}},\;\;\; i=2,4,\ldots,k-1\]  
\item
In the second round, first, quantum player 1 writes on the blackboard
the classical string $A_1$. The other
players read the classical string $A_1$ and proceed to perform the mapping
\[ T: \ket{P_j}\ket{0} \mapsto \ket{P_j} (-1)^{A_j} \ket{0} \]
\item
In the third round, the ``blackboard'' contains the string
$(P_1,A_1)$ and the states
\[ \ket{\chi_i} =
\ket{i}\ket{P_i} (-1)^{A_i} \ket{0} +\ket{i+1}\ket{P_{i+1}} (-1)^{A_{i+1}}
\ket{0}  \;\;\;\mbox{for}\; i=2,4,\ldots,k-1.\]
 
The referee quantumly ``erases'' the inputs resulting in the states  
\[ \ket{\psi_i} = (-1)^{A_i}\ket{i}+ (-1)^{A_{i+1}}\ket{i+1}.\]
By measuring in the basis $\{\ket{i}\pm \ket{i+1}\}$
the referee computes $A_i \oplus A_{i+1}$ exactly. 
\end{itemize}
The correctness of the quantum protocol is the
same as in the classical one, i.e.  $\frac{1}{2}+\delta$.
\end{proof}

Hence, we have proved an exponential separation between randomized and
quantum multiparty communication complexity for a total function.

\section{Conclusions}

We proved a simulation theorem between quantum and classical
multiparty communication complexity. This enabled us to reduce the
question of showing that a function is outside the circuit complexity
class $SYM(\log n, 2^{\omega(polylog n)})$ to the question of finding
an explicit function $f$ for which the $\ell$-party quantum
communication complexity is $\Omega(\frac{n}{2^\ell}+\log \delta)$. 
The main open question is to find such an explicit function. It would
be very interesting to see if the techniques used for proving lower
bounds of the form $\Omega(\frac{n}{2^\ell})$ in the classical case
could be extended in the quantum case. 

Moreover, we showed an exponential separation between classical and
quantum multiparty communication complexity for a total boolean
function. This is the first such separation in any communication model
and leaves open the question of a similar separation in the case of
two-party communication.

\end{document}